\documentclass[12pt]{article}
\usepackage{epsfig}
\usepackage{axodraw2}
\usepackage{a4}
\usepackage{latexsym}
\usepackage{cite}
\usepackage{slashed}
\usepackage{hyperref}
\usepackage{tikz}

\usepackage{color,colordvi}

\newcommand{\gsim}{\raisebox{-0.07cm}{$\:\:\stackrel{>}{{\scriptstyle
 \sim}}\:\: $} }
\newcommand{\lsim}{\raisebox{-0.07cm}{$\:\:\stackrel{<}{{\scriptstyle
 \sim}}\:\: $} }
\newcommand{\equal}{\:\: = \:\:}

\newcommand{\hspn}{{\hspace{-4mm}}}

\newcommand{\beq}{\begin{equation}}
\newcommand{\eeq}{\end{equation}}
\newcommand{\bea}{\begin{eqnarray}}
\newcommand{\eea}{\end{eqnarray}}
\newcommand{\nn}{\nonumber}
\newcommand{\MSb}{$\overline{\mbox{MS}}$}

\newcommand{\als}{\alpha_{\rm s}}

\newcommand{\ep}{\varepsilon}

\begin{document}
\setlength{\parskip}{0.25cm}
\setlength{\baselineskip}{0.56cm}

\def\z#1{{\zeta_{#1}}}
\def\zss{\zeta_2^{\,2}}

\def\nc{{n_c}}
\def\ncs{{n_{c}^{\,2}}}
\def\nct{{n_{c}^{\,3}}}

\def\ca{{C^{}_{\!A}}}
\def\cas{{C^{\,2}_{\!A}}}
\def\cat{{C^{\,3}_{\!A}}}
\def\caf{{C^{\,4}_{\!A}}}
\def\cai{{C^{\,5}_{\!A}}}

\def\cf{{C^{}_F}}
\def\cfs{{C^{\, 2}_F}}
\def\cft{{C^{\, 3}_F}}
\def\cff{{C^{\, 4}_F}}

\def\nf{{n^{}_{\! f}}}
\def\nfz{{n^{\,0}_{\! f}}}
\def\nfo{{n^{\,1}_{\! f}}}
\def\nfs{{n^{\,2}_{\! f}}}
\def\nft{{n^{\,3}_{\! f}}}
\def\nff{{n^{\,4}_{\! f}}}

\def\tf{{T^{}_{\!F}}}
\def\tfs{{T^{\,2}_{\!F}}}
\def\tft{{T^{\,3}_{\!F}}}
\def\tff{{T^{\,4}_{\!F}}}

\def\dfAAna{{\frac{d_A^{\,abcd}d_A^{\,abcd}}{N_A }}} 
\def\dfFAna{{\frac{d_F^{\,abcd}d_A^{\,abcd}}{N_A }}}
\def\dfFFna{{\frac{d_F^{\,abcd}d_F^{\,abcd}}{N_A }}}

\def\as(#1){{\alpha_{\rm s}^{\:#1}}}
\def\ar(#1){{a_{\rm s}^{\:#1}}}

\def\muRs{{\mu_R^{\,2}}}
\def\L{\mathcal{L}}
\def\eps{\epsilon}
\def\dots{..}

\def\frct#1#2{\mbox{\large{$\frac{#1}{#2}$}}}

\newcommand{\ax}[5]{ \tikz[baseline=-2.5pt,inner sep=0pt,outer sep=0pt]{\node[] at (0,0) {\begin{axopicture}(#1,#2)(#3,#4) #5 \end{axopicture}};} }


\begin{titlepage}
\noindent
Nikhef 2017--001 \hfill {\tt arXiv:1701.01404-v2}\\
LTH 1117 \\
\vspace{1.0cm}
\begin{center}
{\LARGE \bf The five-loop beta function of}\\
\vspace{0.3cm}

{\LARGE \bf Yang-Mills theory with fermions}\\ 
\vspace{2.5cm}
\large
F. Herzog$^{\:\!a}$, B. Ruijl$^{\:\! a,b}$, T. Ueda$^{\:\! a}$, 
J.A.M. Vermaseren$^{\:\! a}$ and A. Vogt$^{\:\! c}$\\
\vspace{1.2cm}
\normalsize
{\it $^a$Nikhef Theory Group \\
\vspace{0.1cm}
Science Park 105, 1098 XG Amsterdam, The Netherlands} \\
\vspace{0.5cm}
{\it $^b$Leiden Centre of Data Science, Leiden University \\
\vspace{0.1cm}
Niels Bohrweg 1, 2333 CA Leiden, The Netherlands}\\
\vspace{0.5cm}
{\it $^c$Department of Mathematical Sciences, University of Liverpool\\
\vspace{0.1cm}
Liverpool L69 3BX, United Kingdom}\\
\vspace{3.0cm}
{\large \bf Abstract}
\vspace{-0.2cm}
\end{center}
We have computed the five-loop corrections to the scale dependence of the
renormalized coupling constant for Quantum Chromodynamics (QCD), its
generalization to non-Abelian gauge theories with a simple compact Lie group, 
and for Quantum Electrodynamics (QED). Our analytical result, obtained using 
the background field method, infrared rearrangement via a new diagram-by-diagram
implementation of the R$^*$ operation and the {\sc Forcer} program for 
massless four-loop propagators, confirms the QCD and QED results obtained by 
only one group before. The numerical size of the five-loop corrections is 
briefly discussed in the standard \MSb\ scheme for QCD with $\nf$ flavours and 
for pure SU($N$) Yang-Mills theory. Their effect in QCD is much smaller than 
the four-loop contributions, even at rather low scales.
\vspace*{0.5cm}
\end{titlepage}
\newpage

%
\section{Introduction}

The scale dependence (`running') of the renormalized coupling constant 
$\alpha_{\rm i}^{}$ is a fundamental property of an interacting quantum field 
theory. In renormalization-group improved perturbation theory, the beta 
function governing this dependence can be written as
\beq
\label{betafct}
  \frac{d\:\! a}{d\:\! \ln \mu^2} \;=\; \beta(a)
  \;=\; - \sum_{n=0}^\infty \, \beta_{n}\: a^{n+2}
\; , \quad
  a \:=\: \frac{\alpha_{\rm i}^{}(\mu)}{4\:\!\pi} 
\eeq
where $\mu$ is the renormalization scale. The determination of the 
(sign of the) leading one-loop coefficient $\beta_0$ 
\cite{AF1,AF2,beta0tH,beta0a,beta0b}, 
soon followed by the calculation of the two-loop correction $\beta_1$ 
\cite{beta1a,beta1b}, led to the discovery of the asymptotic freedom of
non-Abelian gauge theories and thus paved the way for establishing QCD
as the theory of the strong interaction.
The~renormalization-scheme dependent three-loop (next-to-next-to-leading 
order, N$^2$LO) and four-loop (next-to-next-to-next-to-leading order, 
N$^3$LO) coefficients $\beta_2$ and $\beta_3$ were computed in 
refs.~\cite{beta2a,beta2b} and \cite{beta3a,beta3b} in minimal subtraction 
schemes \cite{MS,MSbar} of dimensional regularization \cite{DimReg1,DimReg2}.

In the past years, the N$^2$LO accuracy has been reached for many processes 
at high-energy colliders. N$^3$LO corrections have been determined for 
structure functions in inclusive deep-inelastic scattering (DIS) 
\cite{mvvF2L,mvvF3} and for the total cross section for Higgs-boson 
production at hadron colliders \cite{Higgs1,Higgs2}. Some moments of 
coefficient functions for DIS have recently been computed at N$^4$LO 
\cite{avLL2016}. Reaching this order would virtually remove the uncertainty
due to the truncation of the series of massless perturbative QCD in 
determinations of the strong coupling constant $\als$ from the scaling 
violations of structure functions in DIS.

The corresponding five-loop contributions to the beta functions of QCD, 
with all colour factors `hard-wired', and QED have already been computed in 
refs.~\cite{beta4SU3,beta4QED}. Their leading large-$\nf$ contributions 
have long been known \cite{beta4nf4}, and the sub-leading large-$\nf$ terms 
have been checked and generalized to a general simple gauge group in 
ref.~\cite{beta4nf3}. The real tour de force of 
ref.~\cite{beta4SU3} though, are the parts proportional to $\nfz$, $\nfo$ and 
$\nfs$ which together required more than a year of computations on a decent 
number of multi-core workstations in a highly non-trivial theoretical 
framework. These critical parts have neither been extended to a general gauge 
group nor validated by a second independent calculation so far.

In the present article we address this issue and present the five-loop beta 
function for a general simple gauge group. Unlike the calculations in 
refs.~\cite{beta0a,beta0b,beta1a,beta1b,beta1c,beta2a,beta2b,beta3a,beta3b}, 
we have employed the background field method \cite{Abbott80,AbbottGS83}, which 
we found to be more efficient -- in~validation calculations of the 
{\sc Forcer} program \cite{tuACAT2016,tuLL2016,FORCER} of the four-loop 
renormalization of Yang-Mills theories to all powers of the gauge parameter --
than the computation of two propagators and a corresponding vertex. 
This method and other theoretical and calculational issues, in particular a 
new implementation \cite{NEWRSTAR} of the R$^*$ operation 
\cite{RSTAR1982,RSTAR1984,RSTAR1985,RSTAR1991} for massless propagator-like 
diagrams, are addressed in section~2; the details of the required tensor 
reduction can be found in the appendix.
We present and discuss our result in section~3, and briefly summarize our 
findings in section~4. 

%
\section{Theoretical framework and calculations}
\setcounter{equation}{0}
%

In this section we briefly review the background-field formalism and the 
$R^*$ operation. 
We further define our notations for group invariants, and we give an
overview of our calculation.

\subsection{Background field method}
A convenient and efficient method to extract the Yang-Mills beta function 
is to make use of the background field. We will briefly 
review this formalism. A convenient starting point is the Lagrangian 
of Yang-Mills theory coupled to fermions in a non-trivial (often the 
fundamental) representation of the gauge group, the theory for which we 
will present the 5-loop beta-function in the next section.
 
The Lagrangian of this theory can be decomposed as
\beq
\L_{\mathrm{YM+FER}} \;=\;
\L_{\mathrm{CYM}}+\L_{\mathrm{GF}}+\L_{\mathrm{FPG}}+\L_{\mathrm{FER}}\,.
\eeq
Here the classical Yang-Mills Lagrangian (CYM), a gauge-fixing term (GF), 
the Faddeev-Popov ghost term (FPG) and the fermion term (FER) are given by
\bea
\L_{\mathrm{CYM}} &\!=\!& -\frac{1}{4}\, F_{\mu\nu}^a(A)F^{\mu\nu}_a(A)
\, , \nn \\ 
\L_{\mathrm{GF}}  &\!=\!& -\frac{1}{2\xi}\, (G^a)^2
\,, \nn \\[1mm] 
\L_{\mathrm{FPG}} &\!=\!& -\eta^\dagger_a \,\partial^{\:\!\mu} D^{ab}_\mu(A) 
  \,\eta_b^{}
\,, \nn \\[2mm] 
\L_{\mathrm{FER}} &\!=\!& \sum_{i,j,f}^{} \bar \psi_{if}(i\slashed{D}_{ij}(A)
  -m_{\!f}^{}\delta_{ij})\, \psi_{jf}^{}\, .
\eea
In the fermion term the sum goes over colours $i,j$, and $n_f$ flavours $f$, 
and we use the standard Feynman-slash notation. The field strength is given 
by
\beq
\qquad F_{\mu\nu}^a(A) \;=\; \partial_{\:\!\mu} A_\nu^a
  -\partial_{\:\!\nu} A_\mu^a + g f^{abc} A^b_{\mu}A^c_{\nu}
\eeq
and the covariant derivatives are defined as 
\bea
D^{ab}_\mu(A) &\!=\!& \delta^{\:\!ab}\partial_\mu -g f^{abc} A^c_\mu
\, ,  \nn \\[1mm]
D_{ij}^\mu(A) &\!=\!& \delta_{ij}\partial^{\:\!\mu} -ig\, T^{a}_{ij} A_a^\mu  
\, .
\eea
The conventions associated to the generators $T^a$ and structure constants 
$f^{abc}$ of the gauge group will be explained in section~\ref{sec:groups}. 
The gauge-fixing term depends on making a suitable choice for $G^a$, 
which is usually taken as $G^a=\partial^{\:\!\mu} A_\mu^a$. 

The background-field Lagrangian is derived by decomposing the gauge field as
\beq
\label{bfg}
A_\mu^a(x) \;=\; B_\mu^a(x)+\hat A_\mu^a(x)\,,
\eeq
where $B_\mu^a(x)$ is the \emph{classical} background field while $\hat 
A_\mu^a(x)$ contains the \emph{quantum} degrees of freedom of the gauge 
field $A_\mu^a(x)$. The background-field Lagrangian is then written as
\beq
\label{eq:BYM}
\L_{\mathrm{BYM+FER}} \;=\; \L_{\mathrm{BCYM}}+\L_{\mathrm{BGF}}
  +\L_{\mathrm{BFPG}}+\L_{\mathrm{BFER}}\,.
\eeq
$\L_{\mathrm{BCYM}}$ and $\L_{\mathrm{BFER}}$ are derived simply by 
substituting eq.~(\ref{bfg}) into the corresponding terms in the Yang-Mills 
Lagrangian. However a clever choice exists \cite{Abbott80,AbbottGS83} for the 
ghost and gauge fixing terms, which allows this Lagrangian to maintain explicit
gauge invariance for the background field $B_\mu^a(x)$, while fixing only the 
gauge freedom of the quantum field  $\hat A_\mu^a(x)$. The gauge fixing 
then uses instead
\beq
G^a \;=\; D_\mu^{ab}(B) \hat A^\mu_{b}\, ,
\eeq
while the ghost term is given by
\beq
\L_{\mathrm{BFPG}} \;=\;  -\eta^\dagger_a \, D^{ab;\mu}(B) \,
  D^{bc}_\mu(B+\hat A) \,\eta_c\, .
\eeq
The Lagrangian $\L_{\mathrm{BYM+FER}}$ then gives rise to additional  
interactions which are different from the normal QCD interactions of the 
quantum field $\hat A_\mu^a(x)$ also contain interactions of $B_\mu^a(x)$ with 
all other fields. 

\begin{figure}
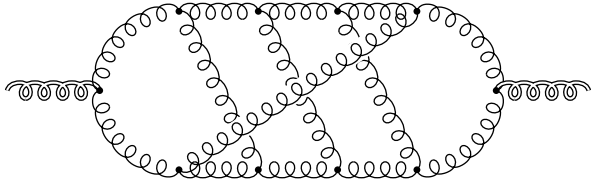

\begin{center}
\begin{axopicture}(230,80)(0,0)
\Gluon(130,70)(160,10){3}{8}
\Gluon(100,70)(130,10){3}{8}
\Gluon(70,70)(100,10){3}{8}
\Line[color=white,width=8](160,70)(70,10)
\DoubleGluon(5,40)(40,40){3}{4}{1.3}
\GluonArc(70,40)(30,90,180){3}{6}
\GluonArc(70,40)(30,180,270){3}{6}
\Gluon(160,70)(70,10){3}{15}
\Gluon(100,70)(70,70){-3}{4}
\Gluon(130,70)(100,70){-3}{4}
\Gluon(160,70)(130,70){-3}{4}
\Gluon(130,10)(160,10){-3}{4}
\Gluon(100,10)(130,10){-3}{4}
\Gluon(70,10)(100,10){-3}{4}
\GluonArc(160,40)(30,270,360){3}{6}
\GluonArc(160,40)(30,0,90){3}{6}
\DoubleGluon(190,40)(225,40){3}{4}{1.3}
\Vertex(40,40){1.3}
\Vertex(190,40){1.3}
\Vertex(70,70){1.3}
\Vertex(100,70){1.3}
\Vertex(130,70){1.3}
\Vertex(160,70){1.3}
\Vertex(70,10){1.3}
\Vertex(100,10){1.3}
\Vertex(130,10){1.3}
\Vertex(160,10){1.3}
\end{axopicture}
\caption{One of the more complicated diagrams.  Single lines represent gluons, 
and the external double lines represent the background field. The presence of 
the 10 purely gluonic vertices creates a large expression after the 
substitution of the Feynman rules.}
\label{fig:gluons}
\end{center}
\end{figure}

A remarkable fact is found when considering the renormalization of this 
Lagrangian. Indeed it turns out, see e.g., \cite{Abbott80,AbbottGS83}, that 
the coupling renormalization, $g\to Z_g\, g$, which determines the beta 
function, is directly related to the renormalization of the background 
field, $B\to B Z_B$, via the identity:
\beq
Z_g\sqrt{Z_B} \;=\;1\, .
\eeq
When working in the Landau gauge, the only anomalous dimension needed in the 
background field gauge formalism is then the beta function. However in the 
Feynman gauge the gauge parameter $\xi$ requires the renormalization 
constant $Z_\xi$ -- which equals the gluon field renormalization constant 
-- but only to one loop lower. In turn this allows one to extract the beta 
function from the single equation
\beq
\label{eq:PIBfinite}
Z_B(1+\Pi_B(Q^2;Z_\xi\xi,Z_g g)) \;=\; \mathrm{finite},
\eeq 
with
\beq
\label{eq:PIBtensor}
\Pi_B^{\mu\nu}(Q;Z_\xi\xi,Z_g g) \;=\;
  (Q^2g^{\mu\nu}-Q^\mu Q^\nu) \: \Pi_B(Q^2;Z_\xi\xi,Z_g g)
\eeq
where $\Pi_B^{\mu\nu}(Q^2;\xi,g)$ is the bare self energy of the background 
field. This self-energy is computed by keeping the fields $B$ external while 
the only propagating fields are $\hat A, \eta$ and $\psi$. A~typical diagram 
which contributes to $\Pi_B(Q^2;\xi,g)$ is given in figure~\ref{fig:gluons}. 

Obtaining the beta function through the background field 
gauge is faster and simpler than the traditional method of 
computing the gluon propagator, ghost propagator and ghost-ghost-gluon 
vertex due to a lower total number of diagrams and the above reduction to a
scalar renormalization.
 
\subsection{Group notations}
\label{sec:groups}
In this section we introduce our notations for the group invariants 
appearing in the results of the next section. $T^{a}$ are the generators of 
the representation of the fermions, and $f^{abc}$ are the structure 
constants of the Lie algebra of a compact simple Lie group, 
\beq
  T^a T^b - T^b T^a \;=\; i f^{abc\,} T^c\, .
\eeq
The quadratic Casimir operators $C_F$ and $C_A$ of the $N$-dimensional 
fermion and the $N_A$-dimensional adjoint representation are given by 
$[T^a T^a]_{ik} \,=\, C_F \delta_{ik}$ and $f^{\,acd} f^{\,bcd} \,=\, 
C_A \delta^{\,ab}$, respectively. The trace normalization of the 
fermion representation is Tr$(T^a T^b) \,=\, T_F \delta^{\,ab}$.
At $L \geq 4$ loops also quartic group invariants enter the beta function. 
These can be expressed in terms of contractions of the totally symmetric 
tensors
\bea
 d_F^{\,abcd} &\! =\! & \frct{1}{6}\: {\rm Tr} ( 
   T^{a\,} T^{b\,} T^{c\,} T^{d\,} + \,\mbox{ five $bcd$ permutations } )
\:\: , \nn \\
 d_A^{\,abcd} &\! =\! & \frct{1}{6}\: {\rm Tr} (  
   C^{a} C^{b} C^{c} C^{d} \:+ \,\mbox{ five $bcd$ permutations } )
\:\: .
\eea
Here the matrices $[\,C^a\,]_{bc} = -i f^{\,abc}$ are the generators of the 
adjoint representation. It should be noted that in QCD-like theories 
without particles that are colour neutral, Furry's theorem~\cite{Furry} 
prevents the occurrence of symmetric tensors with an odd number of indices.

For the fermions transforming according to the fundamental representation
and the standard normalization of the SU($N$) generators, these `colour 
factors' have the values
\bea
\label{colSU(N)}
 &&
 T_F \:=\: \frac{1}{2} 
\; , \quad 
 C_A \:=\: N 
\; , \quad  
 C_F \:=\: \frac{N_A}{2 N} \:=\: \frac{N^2-1}{2 N} 
\; , \quad 
 \dfAAna \:=\: \frac{N^2(N^2+36)}{24}
\; , \nn \\[1mm] && \qquad\quad
 \dfFAna \:=\: \frac{ N(N^2+6)}{48}
\; , \quad
 \dfFFna \:=\: \frac{N^4-6N^2+18}{96\, N^2} 
\; .
\eea
The results for QED (i.e., the group U(1)) are obtained for $C_A=0$, 
$d_A^{\,abcd}=0$, $C_F = 1$, $T_F = 1$, $d_F^{abcd}=1$, and $N_A = 1$. 
For a discussion of other gauge groups the reader is referred to 
ref.~\cite{beta3a}.


\subsection{The $R^*$-operation}
As outlined above, it is possible to extract the five-loop beta function 
from the poles (in the dimensional regulator $\eps$) of the bare background 
field self-energy $\Pi_B(Q)$. At present it is beyond current computational 
capabilities to calculate the required five-loop propagator integrals directly.
The main obstacle preventing such an attempt is the difficulty 
of performing the required integration-by-parts (IBP) reductions. 

\begin{figure}
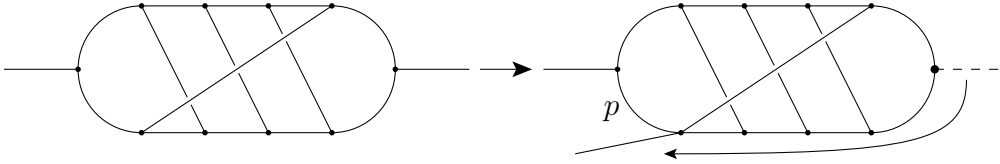

\begin{center}
\begin{axopicture}(390,80)(0,0)
\SetOffset(0,10)
\SetScale{0.8}
\Line(130,70)(160,10)
\Line(100,70)(130,10)
\Line(70,70)(100,10)
\Line[color=white,width=4](160,70)(70,10)
\Line(5,40)(40,40)
\Arc(70,40)(30,90,180)
\Arc(70,40)(30,180,270)
\Line(160,70)(70,10)
\Line(100,70)(70,70)
\Line(130,70)(100,70)
\Line(160,70)(130,70)
\Line(130,10)(160,10)
\Line(100,10)(130,10)
\Line(70,10)(100,10)
\Arc(160,40)(30,270,360)
\Arc(160,40)(30,0,90)
\Line(190,40)(225,40)
\Vertex(40,40){1.3}
\Vertex(190,40){1.3}
\Vertex(70,70){1.3}
\Vertex(100,70){1.3}
\Vertex(130,70){1.3}
\Vertex(160,70){1.3}
\Vertex(70,10){1.3}
\Vertex(100,10){1.3}
\Vertex(130,10){1.3}
\Vertex(160,10){1.3}
\Line[arrow,arrowpos=1,arrowscale=1.5](230,40)(250,40)
\SetOffset(204,10)
\Line(130,70)(160,10)
\Line(100,70)(130,10)
\Line(70,70)(100,10)
\Line[color=white,width=4](160,70)(70,10)
\Line(5,40)(40,40)
\Arc(70,40)(30,90,180)
\Arc(70,40)(30,180,270)
\Line(160,70)(70,10)
\Line(100,70)(70,70)
\Line(130,70)(100,70)
\Line(160,70)(130,70)
\Line(130,10)(160,10)
\Line(100,10)(130,10)
\Line(70,10)(100,10)
\Arc(160,40)(30,270,360)
\Arc(160,40)(30,0,90)
\Line(70,10)(20,0)
\Vertex(40,40){1.3}
\Vertex(190,40){2}
\Vertex(70,70){1.3}
\Vertex(100,70){1.3}
\Vertex(130,70){1.3}
\Vertex(160,70){1.3}
\Vertex(70,10){1.3}
\Vertex(100,10){1.3}
\Vertex(130,10){1.3}
\Vertex(160,10){1.3}
\Text(41,20)[r]{$p$}
\Line[dash,dsize=4](190,40)(220,40)
\Bezier[arrow,arrowpos=1](205,35)(205,5)(180,0)(65,0)
\end{axopicture}
\end{center}
\vspace*{-2mm}
\caption{\small
One external line is moved to create a topology that can be 
integrated. Here we do this for the diagram of figure~\ref{fig:gluons}. One 
should take into account that there can be up to 5 powers of dot products in 
the numerator, causing many subdivergences. Furthermore, the double propagator that 
remains on the right can introduce infrared divergences. After the 
subdivergences have been subtracted, the integral over $p$ can be performed and 
the remaining four-loop topology can be handled by the {\sc Forcer} program.}
\label{fig:rearrange}
\vspace*{5mm}
\end{figure}

Fortunately the problem can be simplified via the use of the $R^*$-operation. 
The $R^*$-operation \cite{RSTAR1982,RSTAR1984,RSTAR1985,RSTAR1991} is a 
subtraction operation capable of rendering any propagator integral finite by 
adding to it a number of suitable subtraction terms. The subtraction terms are 
built from potentially high rank tensor subgraphs of the complete graph, 
whose tensor reduction requires involved methods which we present in appendix~\ref{appendix:tensor}. 
Via the procedure of \textit{IR-rearrangement}, these subtraction terms can 
subsequently be related to simpler propagator integrals.
The IR-rearranged integral is, in general, any other propagator integral 
obtained from the original one by rerouting the external momentum in the 
diagram. This is illustrated in figure~\ref{fig:rearrange}. 

For integrals whose superficial degree of divergence (SDD) is higher than 
logarithmic, the SDD is reduced by differentiating it sufficiently 
many times with respect to its external momenta, before IR-rearranging it.

The upshot of this procedure is that the IR-rearranged propagator integrals 
can be chosen to be \textit{carpet integrals}, which correspond to graphs where 
the external lines are connected only by a single propagator. A carpet integral 
of $L$ loops can be evaluated as a product of an $(L-1)$ loop tensor propagator
 integral times a known one-loop tensor integral. In the case of the five-loop beta 
function this means that we can effectively evaluate the poles of \emph{all} 
five-loop propagator integrals from the knowledge of propagator integrals 
with no more than four loops. A sketch of the $R^*$-operation to compute the superficial
divergence of a 3-loop diagram is shown below:
\begin{eqnarray}
&\Bigg(
\scalebox{0.75}{
\ax{70}{65}{15}{15}{
\Line(12.5,50)(25,50)
\CCirc(50,50){25}{Black}{White}
\Line(75,50)(87.5,50)
\Line(50,25)(50,75)
\CArc(25,25)(25,0,90)
\Vertex(25,50){1.5}
\Vertex(75,50){1.5}
\Vertex(50,25){1.5}
\Vertex(50,75){1.5}
}}
\Bigg)_{\textrm{sup}} = 
\Bigg(
\scalebox{0.75}{
\ax{70}{65}{15}{15}{
\Line(50,75)(50,87.5)
\CCirc(50,50){25}{Black}{White}
\Line(75,50)(87.5,50)
\Line(50,25)(50,75)
\CArc(25,25)(25,0,90)
\Vertex(25,50){1.5}
\Vertex(75,50){1.5}
\Vertex(50,25){1.5}
\Vertex(50,75){1.5}
}
}
\Bigg)_\textrm{sup}
=\nonumber\\
&K \Bigg(
\scalebox{0.75}{
\ax{60}{65}{22}{16}{
\Line(50,75)(50,87.5)
\CCirc(50,50){25}{Black}{White}
\Line(75,50)(87.5,50)
\Line(50,25)(50,75)
\CArc(25,25)(25,0,90)
\Vertex(25,50){1.5}
\Vertex(75,50){1.5}
\Vertex(50,25){1.5}
\Vertex(50,75){1.5}
}
}
-K\Bigg(
\scalebox{0.5}{
\ax{70}{65}{15}{15}{
\Line(12.5,50)(25,50)
\CCirc(50,50){25}{Black}{White}
\Line(75,50)(87.5,50)
\Vertex(25,50){1.5}
\Vertex(75,50){1.5}
}
}
\Bigg)
\scalebox{0.75}{
\ax{60}{65}{22}{16}{
\CCirc(50,50){25}{Black}{White}
\Line(50,75)(50,87.5)
\Line(50,25)(50,75)
\Line(75,50)(87.5,50)
\Vertex(75,50){1.5}
\Vertex(50,25){1.5}
\Vertex(50,75){1.5}
}
}
-
\Bigg(
\scalebox{0.75}{
\ax{50}{65}{20}{15}{
\Line(50,25)(62.5,25)
\Line(50,75)(62.5,75)
\CArc(50,50)(25,90,270)
\Line(50,25)(50,75)
\CArc(25,25)(25,0,90)
\Vertex(25,50){1.5}
\Vertex(50,25){1.5}
\Vertex(50,75){1.5}
}
}
\Bigg)_\textrm{sup}
\scalebox{0.5}{
\ax{70}{65}{15}{15}{
\Line(12.5,50)(25,50)
\CCirc(50,50){25}{Black}{White}
\Line(75,50)(87.5,50)
\Vertex(25,50){1.5}
\Vertex(75,50){1.5}
}
}
\Bigg)
\end{eqnarray}
where \textrm{sup} denotes the superficial divergence, and $K$ isolates the pole of a 
Laurent series in $\eps$. As can be seen, the $R^*$-operation
is recursive, since the same procedure needs to be applied to compute the 
superficial divergence of each counterterm.

The {\sc Forcer} program \cite{tuLL2016,FORCER}, 
written in the {\sc Form} language, is capable to efficiently compute the subtraction terms.
It reduces four-loop propagator integrals to simpler 
known ones by integrating two-point functions, and by applying 
parametrically solved IBP reduction rules to eliminate propagators.
We have automated the $R^*$-operation in a fast {\sc Form} program, 
capable of performing the subtraction of propagator integrals with 
arbitrary tensorial rank. Having interfaced the {\sc Forcer} program with 
the $R^*$ program we were able to compute the poles of all integrals 
entering the five-loop background field self-energy. The algorithms and 
details of our implementation of the $R^*$-operation follow 
to some degree the ideas which were presented in the literature 
 (see e.g., \cite{Caswell,Phi4,RSTAR1982,RSTAR1984,RSTAR1985,RSTAR1991}), 
however we have generalized certain notions in order to deal with arbitrary 
tensor integrals and their associated ultraviolet and infrared divergences. 
These generalizations are subtle and will be presented elsewhere \cite{NEWRSTAR}.


\subsection{Diagram computations and analysis}

The Feynman diagrams for the background propagator up to five loops have been
generated using QGRAF \cite{QGRAF}. They have then been heavily manipulated
by a {\sc Form} \cite{FORM3,TFORM,FORM4} program that determines the topology 
and calculates the colour factor using the program of ref.~\cite{Colour}.
Additionally, it merges diagrams of the same topology, colour factor, 
and maximal power of $\nf$ into meta diagrams for computational efficiency.
Integrals containing massless tadpoles or symmetric colour tensors with an odd
number of indices have been filtered out from the beginning. Lower-order 
self-energy insertions have been treated as described in ref.~\cite{jvLL2016}. 
In this manner we arrive at 2 one-loop, 9 two-loop, 55 three-loop, 
572 four-loop and 9414 five-loop meta diagrams.

The diagrams up to four loops have been computed earlier to all powers of
the gauge parameter using the {\sc Forcer} program 
\cite{tuACAT2016,tuLL2016,FORCER}. For the time being, our five-loop
computation has been restricted to the Feynman gauge, $\xi_F = 1 - \xi = 0$.
An extension to the first power in $\xi_F$ would be considerably slower;
the five-loop computation for a general $\xi$ would be impossible without
substantial further optimizations of our code. 
Instead of by varying~$\xi$, we have checked our computations by verifying 
the relation $\, Q_\mu Q_\nu\, \Pi_B^{\:\!\mu\nu} \,=\, 0\, $ 
required by eq.~(\ref{eq:PIBtensor}). This check took considerably more time
than the actual determination of $\beta_4$.

The five-loop diagrams have been calculated on computers with a combined 
total of more than 500 cores, 80\% of which are older and slower by a factor 
of almost three than the latest workstations. One core of the latter performs 
a `raw-speed' {\sc Form} benchmark, a four-dimensional trace of 14 Dirac 
matrices, in about 0.02 seconds which corresponds to 50 `form units' (fu) 
per hour. The total CPU time for the five-loop diagrams was $3.8 \cdot 
10^{7}$ seconds which corresponds to about $2.6 \cdot 10^{5}$ fu on the 
computers used. The {\sc TForm} parallelization efficiency for single meta 
diagrams run with 8 or 16 cores was roughly 0.5; the whole calculation of $\beta_4$,
distributed `by hand' over the available machines, finished in three days.
 
For comparison, the corresponding $R^*$ computation for $\xi_F = 0$ at
four loops required about $10^{3}$ fu, which is roughly the same as for the
first computation of the four-loop beta function to order $\xi_F^{\,1}$ by a
totally different method in ref.~\cite{beta3a}. The computation with the 
{\sc Forcer} program at four and fewer loops is much faster, in fact fast 
enough to comfortably demonstrate the full three-loop renormalization of 
QCD in 10 minutes on a laptop during a seminar talk \cite{JosSept15}. 

The determination of $Z_B$ from the unrenormalized background
propagator is performed by imposing, order by order, the finiteness of its
renormalized counterpart. The beta function can simply be read
off from the $1/\ep$ coefficients of $Z_B$.
If the calculation is performed in the Landau
gauge, the gauge parameter does not have to be renormalized.
In a $k$-th order expansion about the Feynman gauge
at five loops, the $L\!<5$ loop contributions are needed up to 
$\xi_F^{\,5-L}$. The four-loop renormalization constant for the gauge
parameter is not determined in the background field and has to be `imported'.
In the present $k=0$
case, the terms already specified in ref.~\cite{beta3b} would have been 
sufficient had we not performed the four-loop calculation to all powers of 
$\xi_F$ anyway.


%
\section{Results and discussion}
\setcounter{equation}{0}

Before we present our new result, it may be convenient to recall the beta 
function (\ref{betafct}) up to four loops
\cite{beta0a,beta0b,beta1a,beta1b,beta1c,beta2a,beta2b,beta3a,beta3b}
in terms of the colour factors defined in section~2,
\bea
\label{beta0}
  \beta_{0} &\! =\! &  \frac{11}{3}\: \ca \:-\: \frac{4}{3}\: \tf\, \nf 
\:\: , \\[2mm]
\label{beta1}
  \beta_{1} &\! =\! & \frac{34}{3}\: \cas \:-\: \frac{20}{3}\: \ca\, \tf\, \nf
    - 4\, \cf\, \tf\, \nf 
\:\: , \\[2mm]
\label{beta2}
  \beta_{2} &\! =\! &  \frac{2857}{54}\: \cat 
    \:-\: \frac{1415}{27}\: \cas\, \tf\, \nf
    \:-\: \frac{205}{9}\: \cf\, \ca\, \tf\, \nf \:+\: 2\, \cfs\, \tf\, \nf 
\nn \\[1mm] & & \mbox{\hspn}
    \:+\: \frac{44}{9}\: \cf\, \tfs\, \nfs 
    \:+\: \frac{158}{27}\: \ca\, \tfs\, \nfs  
\:\: , \\[3mm]
\label{beta3}
  \beta_{3} &\! =\! &
    \caf\, \left( \, \frac{150653}{486} - \frac{44}{9}\, \z3 \right)   
    \:+\: \dfAAna\, \left(  - \frac{80}{9} + \frac{704}{3}\,\z3 \right)
\nn \\[1mm] & & \mbox{\hspn}
    \:+\:  \cat\, \tf\, \nf\,  
      \left(  - \frac{39143}{81} + \frac{136}{3}\, \z3 \right)
    \:+\: \cas\, \cf\, \tf\, \nf\, 
      \left( \, \frac{7073}{243} - \frac{656}{9}\,\z3 \right)
\nn \\[1mm] & & \mbox{\hspn}
    \:+\: \ca\, \cfs\, \tf\, \nf\, 
      \left(  - \frac{4204}{27} + \frac{352}{9}\,\z3 \right)
    \:+\: \dfFAna\, \nf\, \left( \, \frac{512}{9} - \frac{1664}{3}\,\z3 \right)
\nn \\[1mm] & & \mbox{\hspn}
    \:+\: 46\, \cft\, \tf\, \nf\, 
    \:+\:  \cas \tfs \nfs 
      \left( \, \frac{7930}{81} + \frac{224}{9}\,\z3 \right)
    \:+\:  \cfs\, \tfs\, \nfs\, 
      \left( \, \frac{1352}{27} - \frac{704}{9}\,\z3 \right)
\nn \\[1mm] & & \mbox{\hspn}
    \:+\:  \ca\, \cf\, \tfs\, \nfs\, 
      \left( \, \frac{17152}{243} + \frac{448}{9}\,\z3 \right)
    \:+\: \dfFFna\, \nfs\, \left( - \frac{704}{9} + \frac{512}{3}\,\z3 \right)
\nn \\[1mm] & & \mbox{\hspn}
    \:+\: \frac{424}{243}\: \ca\, \tft\, \nft\, 
    \:+\: \frac{1232}{243}\: \cf\, \tft\, \nft\,  
\:\: ,
\end{eqnarray} 
where $n_f$ is the number of fermion (in QCD, quark) flavours.
$\beta_n$ are the same in all MS-like schemes \cite{MS,MSbar},
i.e. within the class of renormalization schemes which differ only by a 
shift of the scale $\mu$.
In the same notation and scheme, the five-loop contribution~reads
\bea
\label{beta4}
  \beta_{4} &\! =\! & 
       \cai \left( 
           {8296235 \over 3888} 
         - {1630 \over 81} \*\, \z3 
         + {121 \over 6} \*\, \z4 
         - {1045 \over 9} \*\, \z5 \right) 
\nn \\[1mm] & & \mbox{\hspn}
       \:+\: \dfAAna\, \* \ca \* \, \left( 
         - {514 \over 3}  
         + {18716 \over 3} \,\* \z3 
         - 968 \*\, \z4 
         - {15400 \over 3} \*\, \z5 
           \right) 
\nn \\[1mm] & & \mbox{\hspn}
       \:+\: \caf\* \,\tf\* \,\nf\* \, \left(  
         - {5048959 \over 972}  
         + {10505 \over 81} \* \,\z3 
         - {583 \over 3} \* \,\z4 
         + 1230\* \,\z5 \right)
\nn \\[1mm] & & \mbox{\hspn}
       \:+\: \cat\* \,\cf\* \,\tf\* \,\nf\* \, \left( \,
           {8141995 \over 1944}  
         + 146\* \,\z3 
         + {902 \over 3} \* \,\z4 
         - {8720 \over 3} \* \,\z5 \right)
\nn \\[1mm] & & \mbox{\hspn}
       \:+\: \cas\* \,\cfs\* \,\tf\* \,\nf\* \, \left( 
         - {548732 \over 81}  
         - {50581 \over 27} \* \,\z3 
         - {484 \over 3} \* \,\z4 
         + {12820 \over 3} \* \,\z5 \right)
\nn \\[1mm] & & \mbox{\hspn}
       \:+\: \ca\* \,\cft\* \,\tf\* \,\nf\* \, \left( 
           3717 
         + {5696 \over 3} \* \,\z3 
         - {7480 \over 3} \* \,\z5 \right)
       \:-\: \cff\* \,\tf\* \,\nf\* \, \left( \,
           {4157 \over 6}  
         + 128\* \,\z3 \right)
\nn \\[1mm] & & \mbox{\hspn}
       \:+\: \dfAAna \* \,\tf\* \,\nf\* \, \left( 
           {904 \over 9}  
         - {20752 \over 9} \* \,\z3 
         + 352\* \,\z4 
         + {4000 \over 9} \* \,\z5 \right)
\nn \\[1mm] & & \mbox{\hspn}
       \:+\: \dfFAna \* \,\ca\* \,\nf\* \, \left( \,
           {11312 \over 9}  
         - {127736 \over 9} \* \,\z3 
         + 2288\* \,\z4 
         + {67520 \over 9} \* \,\z5 \right)
\nn \\[1mm] & & \mbox{\hspn}
       \:+\: \dfFAna \* \,\cf\* \,\nf\* \, \left( 
         - 320 
         + {1280 \over 3} \* \,\z3 
         + {6400 \over 3} \* \,\z5 \right)
\nn \\[1mm] & & \mbox{\hspn}
       \:+\: \cat\* \,\tfs\* \,\nfs \* \, \left( \,
           {843067 \over 486}  
         + {18446 \over 27} \* \,\z3 
         - {104 \over 3} \* \,\z4 
         - {2200 \over 3} \* \,\z5 \right)
\nn \\[1mm] & & \mbox{\hspn}
       \:+\: \cas\* \,\cf\* \,\tfs\* \,\nfs \* \, \left( \,
           {5701 \over 162}  
         + {26452 \over 27} \* \,\z3 
         - {944 \over 3} \* \,\z4 
         + {1600 \over 3} \* \,\z5 \right)
\nn \\[1mm] & & \mbox{\hspn}
       \:+\: \cfs\* \,\ca\* \,\tfs\* \,\nfs \* \, \left( \,
           {31583 \over 18}  
         - {28628 \over 27} \* \,\z3 
         + {1144 \over 3} \* \,\z4 
         - {4400 \over 3} \* \,\z5 \right)
\nn \\[1mm] & & \mbox{\hspn}
       \:+\: \cft\* \,\tfs\* \,\nfs \* \, \left( 
         - {5018 \over 9}  
         - {2144 \over 3} \* \,\z3 
         + {4640 \over 3} \* \,\z5 \right)
\nn \\[1mm] & & \mbox{\hspn}
       \:+\: \dfFAna \* \,\tf\* \,\nfs \* \, \left( 
         - {3680 \over 9}  
         + {40160 \over 9} \* \,\z3 
         - 832\* \,\z4 
         - {1280 \over 9} \* \,\z5 \right)
\nn \\[1mm] & & \mbox{\hspn}
       \:+\: \dfFFna \* \,\ca\* \,\nfs \* \, \left( 
         - {7184 \over 3}  
         + {40336 \over 9} \* \,\z3 
         - 704\* \,\z4 
         + {2240 \over 9} \* \,\z5 \right)
\nn \\[1mm] & & \mbox{\hspn}
       \:+\: \dfFFna \* \,\cf\* \,\nfs \* \, \left( \,
           {4160 \over 3}  
         + {5120 \over 3} \* \,\z3 
         - {12800 \over 3} \* \,\z5 \right)
\nn \\[1mm] & & \mbox{\hspn}
       \:+\: \cas\* \,\tft\* \,\nft \* \, \left( 
         - {2077 \over 27}  
         - {9736 \over 81} \* \,\z3 
         + {112 \over 3} \* \,\z4 
         + {320 \over 9} \* \,\z5 \right)
\nn \\[1mm] & & \mbox{\hspn}
       \:+\: \ca\* \,\cf\* \,\tft\* \,\nft \* \, \left(  
         - {736 \over 81}  
         - {5680 \over 27} \* \,\z3 
         + {224 \over 3} \* \,\z4 \right)
\nn \\[1mm] & & \mbox{\hspn}
       \:+\: \cfs\* \,\tft\* \,\nft \* \, \left( 
         - {9922 \over 81}  
         + {7616 \over 27} \* \,\z3 
         - {352 \over 3} \* \,\z4 \right)
\nn \\[1mm] & & \mbox{\hspn}
       \:+\: \dfFFna \* \,\tf\* \,\nft\* \, \left( \,
           {3520 \over 9}  
         - {2624 \over 3} \* \,\z3
         + 256\* \,\z4 
         + {1280 \over 3} \* \,\z5 \right)
\nn \\[1mm] & & \mbox{\hspn}
       \:+\: \ca\* \,\tff\* \,\nff \* \, \left( \,
           {916 \over 243}  
         - {640 \over 81} \* \,\z3 \right)
       \:-\: \cf\* \,\tff\* \,\nff \* \, \left( \,
           {856 \over 243}  
         + {128 \over 27} \* \,\z3 \right)
\:\: .
\eea
$\zeta$ denotes the Riemann zeta function with $\z3 \,\cong\, 1.202056903$,
$\z4 = \pi^4/90 \,\cong\, 1.08232323$ and $\z5 \,\cong\, 1.036927755$. 
As expected from the lower-order and QED results, higher values of the zeta 
function do not occur despite their occurrence in the results for individual 
diagrams; for further discussions see ref.~\cite{beta4QED,masters4}.

Inserting the group factors of SU($3$) as given in eq.~(\ref{colSU(N)}) leads
to the {\bf QCD} results
\bea
  \beta_{0} &\! =\! &  
         11
   \:-\: {2 \over 3} \,\* \nf
\:\: , \qquad
  \beta_{1} \;=\;
         102
   \:-\: {38 \over 3} \,\* \nf
\:\: , \nn \\[1mm]
  \beta_{2} &\! =\! &  
         {2857 \over 2}
   \:-\: {5033 \over 18} \,\* \nf
   \:+\: {325 \over 54} \,\* \nfs
\:\: , \nn \\[2mm]
\label{b0123qcd}
  \beta_{3} &\! =\! &
         {149753 \over 6} + 3564 \,\* \z3
   \:+\: \nf \,\* \left( -{1078361 \over 162} - {6508 \over 27}\,\* \z3 \right)
\nn \\[1mm] & & \mbox{\hspn}
   \:+\: \nfs\,\* \left( \,{50065 \over 162} + {6472 \over 81} \,\* \z3 \right)
   \:+\: {1093 \over 729} \,\* \nft
\eea 
and
\bea
\label{b4qcd}
  \beta_{4} &\! =\! &
   {8157455 \over 16} + {621885 \over 2} \,\* \z3
     \:-\: {88209 \over 2} \,\* \z4 - 288090 \,\* \z5
\nn \\[1mm] & & \mbox{\hspn\hspn}
   \:+\: \nf \,\* \left( - {336460813 \over 1944} - {4811164 \over 81} \,\* \z3 
     + {33935 \over 6} \,\* \z4 + {1358995 \over 27} \,\* \z5 \right)
\nn \\[1mm] & & \mbox{\hspn\hspn}
   \:+\: \nfs \,\* \left( \, {25960913 \over 1944} + {698531 \over 81} \,\* \z3
     - {10526 \over 9} \,\* \z4 - {381760 \over 81} \,\* \z5 \right)
\nn \\[1mm] & & \mbox{\hspn\hspn}
   \:+\: \nft  \,\*  \left(  - {630559 \over 5832} - {48722 \over 243} \,\* \z3
     + {1618 \over 27} \,\* \z4 + {460 \over 9} \,\* \z5 \right)
   \:+\: \nff \,\* \left( \, {1205 \over 2916} - {152 \over 81} \,\* \z3 \right)
\:\: .\quad
\eea
%
In truncated numerical form $\beta_3$ and $\beta_4$ are given by
\bea
\label{bs3num}
  \beta_{3} &\! \cong\! &
       29242.964
     - 6946.2896 \*\, \nf 
     + 405.08904 \*\, \nfs 
     + 1.499314  \*\, \nft
\; , \quad\\[1mm] 
\label{bs4num}
  \beta_{4} &\! \cong\! &
       537147.67
     - 186161.95 \*\, \nf
     + 17567.758 \*\, \nfs
     - 231.2777 \*\, \nft
     - 1.842474 \*\, \nff
\; . \quad
\eea
In contrast to $\beta_0$, $\beta_1$, and $\beta_2$, which change sign at about
$\nf = 16.5$, 8.05, and 5.84 respectively, $\beta_3$ and $\beta_4$ are positive
(except at very large $\nf$ for $\beta_4$), but have a (local) minimum at 
$\nf \simeq 8.20$ and $\nf \simeq 6.07$.

The corresponding analytical result for {\bf QED}, in the same renormalization
scheme(s) but defined without the overall minus sign in eq.~(\ref{betafct}) is 
given by
\bea
  \beta_{0} &\! =\! &
    {4 \over 3} \,\* \nf
\:\: , \quad
  \beta_{1} \;=\;
    4 \,\* \nf
\:\: , \quad
  \beta_{2} \;=\;
   \:-\: 2 \,\* \nf
   \:-\: {44 \over 9} \,\* \nfs  
\:\: , \nn \\[2mm]
\label{b0123qed}
  \beta_{3} &\! =\! &
   \:-\: 46 \,\* \nf
   \:+\: \nfs \,\* \left( \, {760 \over 27} - {832 \over 9} \,\* \z3 \right)
   \:-\: {1232 \over 243} \,\* \nft
\eea
and
\bea
\label{b4qed}
  \beta_{4} &\! =\! &
         \nf \,\* \left( \, {4157 \over 6} + 128 \,\* \z3 \right)
   \:+\: \nfs \,\* \left( -{7462 \over 9} - 992 \,\* \z3 + 2720\,\* \z5 \right)
\nn \\[1mm] & & \mbox{\hspn\hspn}
   \:+\: \nft \,\* \left(  - {21758 \over 81} + {16000 \over 27} \,\* \z3
     - {416 \over 3} \,\* \z4 - {1280 \over 3} \,\* \z5 \right)
   \:+\: \nff \,\* \left( \, {856 \over 243} + {128 \over 27} \,\* \z3 \right)
\:\: .\quad
\eea
The (corresponding parts of the) results (\ref{beta4}), (\ref{b4qcd}) and
(\ref{b4qed}) are in complete agreement with the findings of
refs.~\cite{beta4SU3,beta4QED,beta4nf4,beta4nf3}.
Consequently, eq.~(\ref{b4qed}) also agrees with the result for QED at $\nf=1$, 
which was obtained in ref.~\cite{QEDnf1} somewhat earlier than the general 
result \cite{beta4QED}.

As already noted in ref.~\cite{beta4SU3}, the five-loop {\bf QCD} coefficient 
of the beta function is rather small [$\:\!$recall that we use a convenient but 
very small expansion parameter in eq.~(\ref{betafct})]. 
Indeed, for the physically relevant values of $\nf$ the expansion in powers
of $\als$ reads
\bea
\label{bSnf3}
  \widetilde{\beta}(\als,\nf\!=\!3) &\! =\! &
       1
       + 0.565884 \,\* \als
       + 0.453014 \,\* \as(2)
       + 0.676967 \,\* \as(3)
       + 0.580928 \,\* \as(4)
\; , \quad \nn \\[1mm]
  \widetilde{\beta}(\als,\nf\!=\!4) &\! =\! &
       1
       + 0.490197 \,\* \als
       + 0.308790 \,\* \as(2)
       + 0.485901 \,\* \as(3)
       + 0.280601 \,\* \as(4)
\; , \quad \nn \\[1mm]
  \widetilde{\beta}(\als,\nf\!=\!5) &\! =\! &
       1
       + 0.401347 \,\* \als
       + 0.149427 \,\* \as(2)
       + 0.317223 \,\* \as(3)
       + 0.080921 \,\* \as(4)
\; , \quad \nn \\[1mm]
  \widetilde{\beta}(\als,\nf\!=\!6) &\! =\! &
       1
       + 0.295573 \,\* \als
       - 0.029401 \,\* \as(2)
       + 0.177980 \,\* \as(3)
       + 0.001555 \,\* \as(4)
\; ,
\eea
where $\widetilde{\beta} \equiv - \beta(a_{\rm s}) / (\ar(2) \beta_0)$ 
has been re-expanded in powers of $\als = 4 \pi\, a_{\rm s}$.
Clearly there is no sign so far of a possible divergence of the perturbation 
series for this quantity.

In order to further illustrate the $\nf$-dependent convergence (or the lack
thereof) of the beta function of QCD, we introduce the quantity
\beq
\label{eq:ashat}
  \widehat{\alpha}_{\rm s}^{\,(n)}(\nf) \; = \; 4 \pi\,
  \left|\, \frac{\beta_{n-1}(\nf)}{4\,\beta_{n}(\nf)} \,\right| \:\: .
\eeq
Recalling the normalization (\ref{betafct}) of our expansion parameter, 
$\widehat{\alpha}_{\rm s}^{\,(n)}(\nf)$ represents the value of $\als$ for 
which the $n$-th order correction is 1/4 of that of the previous order. 
Therefore, $\als \lsim \widehat {\alpha}_{\rm s}^{\,(n)}(\nf)$ defines 
(somewhat arbitrarily due to the choice of a factor of 1/4) a region of fast 
convergence of $\beta(\als,\nf)$.
Obviously, the absolute size of the $n$-th and $(n\!-\!1)$-th order
effects are equal for $\als= 4\,\widehat{\alpha}^{\,(n)}(\nf)$. Thus the
quantity (\ref{eq:ashat}) also indicates where the expansion appears
not to be reliable anymore, 
$\als \gsim 4\,\widehat{\alpha}_{\rm s}^{\,(n)}(\nf)$, 
for a given value of $\nf$ that is not too close to zeros or minima of the 
coefficients $\beta_{n-1}$ and $\beta_n$.


It is interesting to briefly study the $N$-dependence of the convergence 
behaviour for the case of SU($N$) gauge theories. For our brief illustration
we confine ourselves to pure Yang-Mills theory, $\nf = 0$, and consider 
\beq
\label{eq:asunhat}
  \widehat{\alpha}_{\rm YM}^{\,(n)}(N) \; = \; 4 \pi\, N 
  \left|\, \frac{\beta_{n-1}(N)}{4\,\beta_{n}(N)} \,\right| \:\: ,
\eeq
where the factor $N$ compensates the leading large-$N$ dependence $N^{n+1}$
of $\beta_n$, i.e., the parameter that needs to be small in SU($N$) Yang-Mills 
theory is not $\alpha_{\rm YM}^{}$ but $N \alpha_{\rm YM\,}^{}$.

\begin{figure}[p]
\vspace{-4mm}
\centerline{\epsfig{file=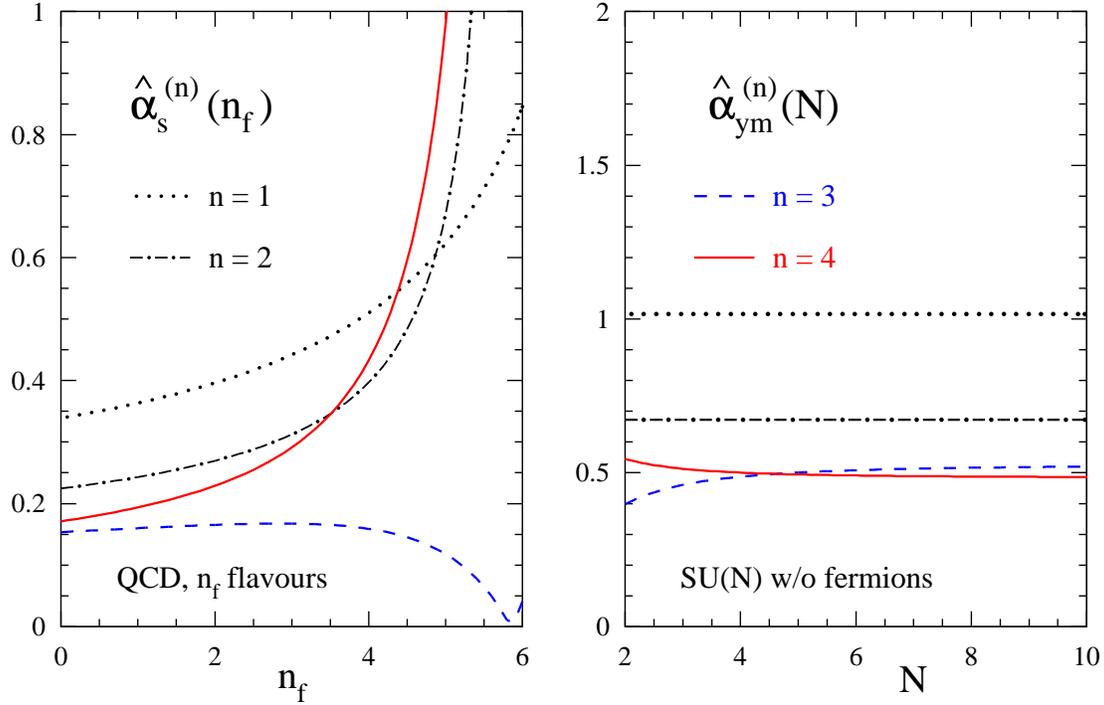,width=15.0cm,angle=0}}
\vspace{-2mm}
\caption{ \label{fig:ashat} \small
 The values (\ref{eq:ashat}) and (\ref{eq:asunhat}) of the coupling constants
 of QCD (left) and pure SU($N$) Yang-Mills theory (right) for which the absolute
 size of the N$^n$LO contribution to the beta function is a quarter of that of 
 the N$^{n-1}$LO term for $n = 1$, 2, 3 (dashed curves) and 4 (solid curves).}
\vspace{-1mm}
\end{figure}

The quantities (\ref{eq:ashat}) and (\ref{eq:asunhat}) are displayed in the 
left and right panel of figure~\ref{fig:ashat}, respectively. The behaviour of
$\widehat{\alpha}_{\rm s}^{\,(n)}$ at the upper end of the $\nf$ range 
shown in the figure is affected by the zeros and minima of the coefficients
$\beta_n>0$ mentioned below eq.~(\ref{bs4num}). 
The $N$-dependence of $\widehat{\alpha}_{\rm YM}$ for pure Yang-Mills theory, 
where only terms with $N^{n+1}$ and $N^{n-1}$ enter $\beta_{n}$ (the latter 
only at $n \geq 4$ via 
$d_A^{\,abcd}d_A^{\,abcd}/N_A$, cf.~eq.~(\ref{colSU(N)}) above), is rather 
weak. With only the curves up to four loops, one might be tempted to draw
conclusions from the shrinking of the `stable' $\als$ region from NLO to
N$^2$LO and from N$^2$LO to N$^3$LO that are not supported by the N$^4$LO 
(five-loop) results of ref.~\cite{beta4SU3} and the present article.

\begin{figure}[p]
\vspace{-2mm}
\centerline{\epsfig{file=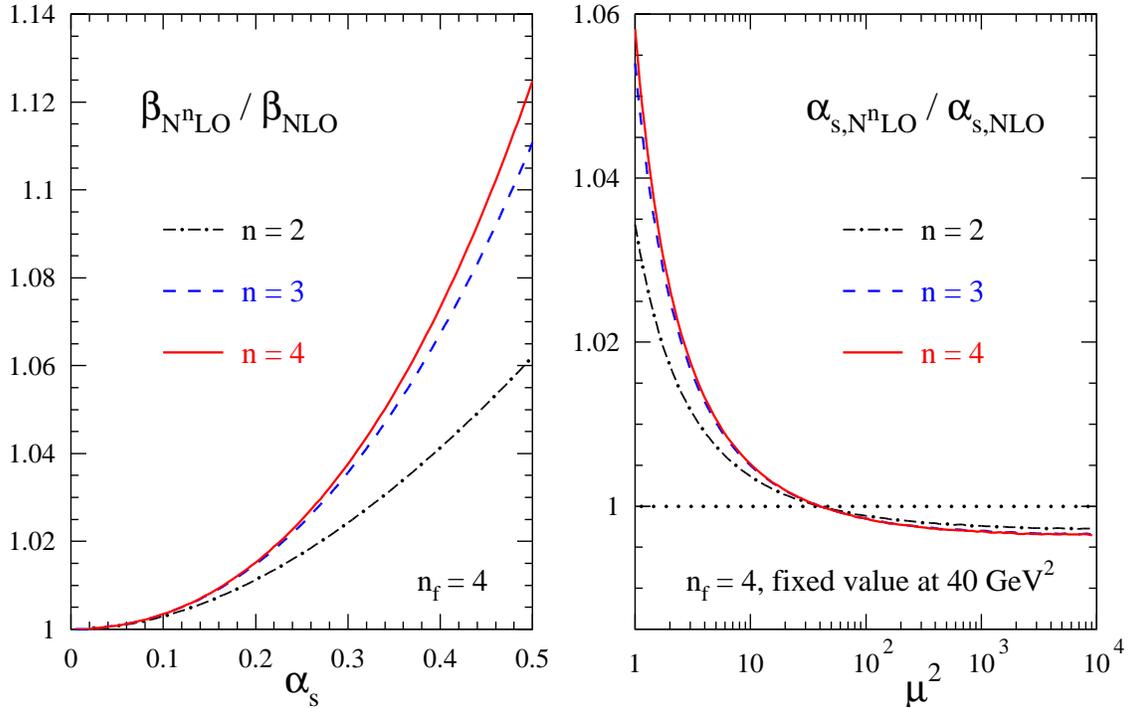,width=15.0cm,angle=0}}
\vspace{-2mm}
\caption{ \label{fig:asrun} \small
 Left panel: The total N$^2$LO, N$^3$LO and N$^4$LO results for the beta 
 function of QCD for four flavours, normalized to the NLO approximation.
 Right panel: The resulting scale dependence of $\als$ for a value of 0.2 at 
 $40 \mbox{ GeV}^2$, also normalized to the NLO result in order to show the 
 small higher-order effects more clearly, for the scale range 
 $1 \mbox{ GeV}^2 \leq \mu^2 \leq 10^{\,4} \mbox{ GeV}^2$.}
\vspace{-1mm}
\end{figure}

Finally, we briefly illustrate the cumulative effect of the orders up to N$^4$LO
on the beta function of QCD and the scale dependence of the strong coupling
constant $\als$ in figure~\ref{fig:asrun}. 
For this illustration we set $\nf = 4$ and choose, in order to only show the 
differences caused by the running of the coupling, an order-independent value 
of $\als = 0.2$ at $\mu^2 = 40 \mbox{ GeV}^2$.  A realistic order dependence 
of $\als$ at this scale, as determined from the scaling violations in DIS, 
would be 0.208, 0.201, 0.200, and 0.200 at NLO, N$^2$LO, N$^3$LO, and N$^4$LO, 
respectively \cite{mvvF2L}.
 
Adding the N$^4$LO contributions changes the beta function by less than 1\%
at $\als = 0.47$ for $\nf = 4$ and at $\als = 0.39$ for $\nf = 3$; the 
corresponding values at N$^3$LO are 0.29 and 0.26. The N$^4$LO effect on the
values of $\als$ as shown in figure~\ref{fig:asrun} are as small as 0.08\%
(0.4\%) at $\mu^2 = 3 \mbox{ GeV}^2$ ($1 \mbox{ GeV}^2$); the corresponding
N$^3$LO corrections are 0.5\% (2\%). Of course these results do not preclude
sizeable purely non-perturbative corrections, but it appears that the 
perturbative running of $\als$ is now fully under control for all practical 
purposes. 



%
\section{Summary and outlook}
%

The five-loop (next-to-next-to-next-to-next-to-leading order, N$^4$LO)
coefficient $\beta_4$ of the renormalization-group beta function has been 
computed in MS-like schemes for Yang-Mills theories with a simple compact 
Lie group and one set of $\nf$ spin-1/2 fermions. 
This computation confirms and extends the QCD and QED results first 
obtained, respectively, in ref.~\cite{beta4SU3} -- where also some direct 
phenomenological applications to $\als$ determinations from, e.g., 
$\tau$-lepton decays and Higgs-boson decay have already been discussed -- 
and ref.~\cite{beta4QED}. It~also agrees with the high-$\nf$ partial 
results of refs.~\cite{beta4nf4,beta4nf3}. 

We have illustrated the size of the resulting N$^4$LO corrections to the scale 
dependence of the coupling constant for $\als$-values relevant to \MSb, the
default scheme for higher-order calculations and analyses in perturbative QCD. 
For physical values of $\nf$, the N$^4$LO corrections to the beta function 
are much smaller than the N$^3$LO contributions and amount to 1\% or less, 
even for $\als$-values as large as 0.4. More generally, there is no evidence 
of any increase of the coefficients indicative of a non-convergent perturbative
expansion for the beta functions of QCD and SU($N$) gauge theories.

Our computation has been made possible by the development of a refined 
algorithm \cite{NEWRSTAR}, implemented in {\sc Form} \cite{FORM3,TFORM,FORM4}, 
for the determination of the ultraviolet and infrared divergences of arbitrary 
tensor self-energy integrals via the R$^*$~operation 
\cite{RSTAR1982,RSTAR1984,RSTAR1985,RSTAR1991} 
--- for another recent diagrammatic implementation of R$^*$ for scalar 
integrals and its application to $\varphi^{\,4}$ theory at six loops, see 
refs.~\cite{Batkovich1,Batkovich2} ---
and the {\sc Forcer} program \cite{tuACAT2016,tuLL2016,FORCER} for the 
parametric reduction of four-loop self-energy integrals.
It should be noted that this approach is quite different from those taken in
refs.~\cite{beta4SU3} and \cite{beta4nf3}. In the former the R$^*$~operation
has been carried out `globally', the latter uses a five-loop extension of the
method of fully massive vacuum diagrams as applied for the determination 
of the four-loop beta function in refs.~\cite{beta3a,beta3b}; see also 
ref.~\cite{LMMSqq5}. 
 
One may expect that the present implementation of the R$^*$~operation will be 
useful for other multi-loop calculations, at least after further optimizations. 
An example is the computation of the fifth-order contributions to the anomalous 
dimensions of twist-2 spin-$N$ operators in the light-cone operator product 
expansion, which now represent the only missing piece for full N$^4$LO
analyses of low-$N$ moments of the structure functions $F_2$ and $F_3$
in inclusive deep-inelastic scattering.

A {\sc Form} file with our result for the coefficient $\beta_4$ and its
lower-order counterparts can be obtained from the
preprint server \ {\tt http://arXiv.org} by downloading the source of this 
article. It will also be available from the authors upon request.

%
\vspace*{-2mm}
\subsection*{Acknowledgements}
\vspace*{-1mm}
We would like to thank K. Chetyrkin and E. Panzer for useful discussions.
This work has been supported by the {\it European Research Council}$\,$ (ERC) 
Advanced Grant 320651, {\it HEPGAME} and the UK {\it Science \& Technology
Facilities Council}$\,$ (STFC) grant ST/L000431/1. 
We also are grateful for the opportunity to use most of the {\tt ulgqcd} 
computer cluster in Liverpool which was funded by the STFC grant ST/H008837/1
and to S. Downing for the administration of this now eight year old facility. 

\appendix
%
\section{Tensor reduction}
\label{appendix:tensor}
\setcounter{equation}{0}
%
%
It can be shown that the tensor reduction of ultraviolet and infrared 
subtraction terms, required for the $R^*$-operation, is equivalent
to the tensor reduction of tensor vacuum bubble integrals. In general 
tensor vacuum integrals can be reduced to linear combinations of products of 
metric tensors $g^{\mu\nu}$ whose coefficients are scalar vacuum integrals.
Specifically a rank $r$ tensor, $T^{\mu_1\dots\,\mu_r}$, is written as a 
linear combination of $n=r!/2^{(r/2)}/(r/2)!$ combinations 
of $(r/2)$ metric tensors with coefficients $c_\sigma$, i.e.,
\beq
T^{\mu_1\dots\,\mu_r}=\sum_{\sigma \in\, {}_2 S_{r}} c_\sigma 
 \,T^{\mu_1\dots\mu_r}_\sigma\,,\qquad
T^{\mu_1\dots\,\mu_r}_\sigma= g^{\mu_{\sigma(1)}\mu_{\sigma(2)}} 
\dots\, g^{\mu_{\sigma(r-1)}\mu_{\sigma(r)}}\,.
\eeq
Here we define ${}_2S_{r}$ as the set of permutations which do \emph{not} 
leave the tensor $T^{\mu_1\dots\,\mu_r}_\sigma$ invariant. The coefficients 
$c_\sigma$ can be obtained by acting onto the tensor $T^{\mu_1\dots\,\mu_r}$ 
with certain projectors $P_\sigma^{\mu_1\dots\mu_r}$, such that
\beq
c_\sigma=P_\sigma^{\:\!\mu_1\dots\,\mu_r} T_{\mu_1\dots\,\mu_r}\,.
\eeq
From this it follows that the orthogonality relation,
\beq
\label{eq:orthogonality}
 P_\sigma^{\:\!\mu_1\dots\,\mu_r} T_{\tau,\,\mu_1\dots\,\mu_r} 
 = \delta_{\sigma\tau}\,,
\eeq
must hold, where $\delta$ is the Kronecker-delta. Since the projector 
$P_\sigma^{\:\!\mu_1\dots\,\mu_r}$ of each tensor can also be written in terms 
of a linear combination of products of metric tensors, inverting an $n\times n$
matrix determines all the projectors. However, the size of the matrix grows 
rather rapidly as $r$ increases.  Instead of solving an $n \times n$ linear 
system, the symmetry group of the metric tensors can be utilized to reduce the 
size of the system.
From eq.~(\ref{eq:orthogonality}) it follows that the projector $P_\sigma$ is in the 
same symmetry group (the group of 
permutations which leave it invariant) as $T_\sigma$.
For example, given a permutation $\sigma_1=(123...(r-1)r)$,
\beq
T_{\sigma_1}^{\mu_1\dots\,\mu_r} = g^{\mu_1\mu_2} g^{\mu_3\mu_4} 
 \dots\, g^{\mu_{r-1}\mu_r}\,.
\eeq
The corresponding projector $P_{\sigma_1}^{\mu_1\dots\,\mu_r}$ must be 
symmetric under interchanges of indices such as $\mu_1 \leftrightarrow 
\mu_2$, $(\mu_1,\mu_2) \leftrightarrow (\mu_3,\mu_4)$ and so on.
Grouping the metric tensors by the symmetry leads to the fact that 
$P_\sigma$ is actually written in a linear combination of a small number of 
$m$ tensors instead of $n$ ($m \le n$),
\beq
P_\sigma^{\mu_1\dots\,\mu_r} =\sum_{k=1}^m b_k 
 \sum_{\tau\in A_m^\sigma} T^{\mu_1\dots\,\mu_r}_\tau.
\eeq
The $m$ sets of permutations $A_{k=1\dots m}^\sigma$ must therefore each be closed 
under the permutations which leaves $T_\sigma$ invariant and at the same 
time their union must cover once the set ${}_2S_n$. 
Contracting $P_\sigma$ with $T_\tau$s where we choose a representative permutation $\tau$ 
from each $A_{k}^\sigma$, i.e one permutation from $A_1^\sigma$, one permutation from $A_2^\sigma$ etc, gives an $m \times m$ matrix which can be inverted 
to yield the coefficients $b_k$. The number of unknowns $m$ is, for example $m=5$ for 
$r=8$ and $m=22$ for $r=16$, which are compared to $n=105$ for $r=8$ and $n=2027025$ for $r=16$. 
The comparison of these numbers illustrates that the exploitation of the 
symmetry of the projectors makes it possible to find the tensor reduction 
even for very large values of~$r$, which could never have been obtained 
by solving the $n\times n$ matrix.

%

{
\setlength{\baselineskip}{0.54cm}

}
\end{document}